\documentclass{cimento}

\usepackage{graphicx}

\newenvironment{packed_itemize}{
\begin{itemize}
     }{\end{itemize}}

\def\eg{{e.g.}}
\newcommand{\pt}       {\ensuremath{p_{T}}}

\newcommand{\invfb}{\ensuremath{\mathrm{fb^{-1}}}}
\newcommand{\ttbar}  {\ensuremath{ t\bar{t}}}
\newcommand{\GeV}{\ensuremath{\mathrm{GeV}}}
\newcommand{\tev}{\ensuremath{\mathrm{TeV}}}
\newcommand{\gev}{\GeV}

\newcommand{\ppbar}{\ensuremath{p\bar p}}

\title{Recent results from D0  and recent Tevatron combinations}
\author{B.~Tuchming   \from{ins:x}
  \thanks{ on behalf of the CDF and D0 collaborations} }
\instlist{ \inst{ins:x}     CEA Saclay, Irfu, SPP, F-91191 Gif-Sur-Yvette Cedex, France}

\begin{document}

\maketitle

\begin{abstract}
  We discuss recent results obtained by the D0 experiment at the Fermilab Tevatron $p\bar p$ collider and recent combinations of  CDF and D0 measurements.
  Regarding the top quark mass, we present recent measurements obtained at D0, the final Run~I+Run~II D0 combination, as well as the preliminary Run~I+Run~II CDF+D0 combination.
  We also discuss the combination of the CDF+D0 measurements of the $p\bar p  \to t\bar t$
  forward-backward asymmetry.
  We present the first measurement of the direct CP-violating charge asymmetry in $B^\pm$ mesons decaying to $\mu^\pm D^{0} X$   and discuss    the status of the recent evidence for a four-flavor tetraquark state seen at D0.

\end{abstract}

\section{Introduction}
The Fermilab Tevatron $p\bar p$  collider has been finally shut down in 2011,  after many years of successful running.
During  Run~I (1992--1996) at $\sqrt s=1.8~\tev$ and  Run~II (2001--2011) at  $\sqrt s=1.96~\tev$,
it provided approximately
 0.1~\invfb\ and 10~\invfb\ , respectively, of collision data to both Tevatron experiments, CDF and D0.
  The analysis of Tevatron data is not finished and new physics results are still obtained nowadays.
We present here the  most recent results obtained by the D0 experiment and the recent combinations of the CDF and D0 measurements. These new results 
concern top quark physics, heavy flavor physics, and hadron spectroscopy.

\section{Top quark mass}
The top quark  is the heaviest known fundamental particle of the Standard Model (SM) with
 a mass of~$\simeq 175$~GeV close to the  electroweak breaking scale.
 Precise measurements of its mass allow for consistency checks of the SM and could also determine whether the electroweak vacuum is unstable,  metastable, or stable~\cite{Degrassi:2012ry}.

 Within the SM, the top quark decays into a  $W$ boson and a $b$ quark
 almost 100\% of the time. The different $p\bar p\to t\bar t$ channels arise from the possible decays of the pair of $W$ bosons.
 To measure the top quark mass D0 and CDF  perform  various analyses, using the three different channels,  lepton+jets, dilepton, and fully hadronic, and employing different methodologies.  The methodologies can be classified in three  families:
\begin{packed_itemize}
\item
  The template methods are  based on  kinematic observables sensitive to the top quark mass.
Templates of their distributions are built, using Monte Carlo simulation (MC) for $\ttbar$ signal at different $m_ {top}$.
A likelihood fit to the data determines the best signal template.
\item 
The matrix element method (ME) is based on an event probability computed from  the
proton parton distribution functions (PDF), the matrix elements for  the $q\bar q \to t\bar t$ processes, and the transfer functions which relate detector measurements (\eg:  \pt\ of jets)
to the  top quark decay products (\eg:  \pt\ of partons).
Unmeasured quantities or quantities suffering from significant measurement uncertainty are integrated out. A maximum likelihood fit determines the measured  mass. This method needs to be calibrated using MC events.
\item The cross-section method derives the top quark  mass from the measured $\ttbar$ production cross-section. This method measures the well defined pole mass of the top quark, as opposed to the ``MC mass'' measured by the first two methods. The difference between the pole and MC mass was estimated to $O(1)$~GeV for  many years.
  In a  recent work, a difference of $+0.6\pm0.3$~GeV between the MC mass and the pole mass is obtained in the context of an $e^+e^- \to \ttbar$  simulation~\cite{Butenschoen:2016lpz}. However,   further studies would be needed to produce a similar estimate  in the  context of  $\ppbar$ collisions.
\end{packed_itemize}

\subsection{Direct measurements of the top quark mass at D0}
D0 has recently measured the top quark mass in the dilepton channel using the full Run~II dataset of 9.7~\invfb\ and the ME method~\cite{Mtop2-D0-di-l-ME-PRD}. This measurement
uses the jet energy scale (JES) constraint arising from the calibration procedure performed in the lepton+jets channel~\cite{Mtop2-D0-l+jt-PRL,Mtop2-D0-l+jt-PRD}, which reduces the dominant JES uncertainty by a factor of $\approx 3$. The calibration exploits the $W\to q\bar{ q^{\prime}}$
decay of the lepton+jets channel by requiring the mass of the corresponding dijet system to be consistent
with the mass of the $W$ boson ($80.4$~\gev).
The dilepton ME measurement achieves a 1.1\% uncertainty: $ m_t = 173.93\pm 1.61  \mbox{ (stat)}   \pm  0.88\mbox{ (syst)}~\gev
$. Another  dilepton measurement using templates and neutrino-weighting (NW)
technique was published earlier in 2016~\cite{Mtop2-D0-di-l-Nu-PLB}. 
The NW measurements is $ m_t =173.32 \pm 1.32  \mbox{ (stat)}  \pm 0.85 \mbox{ (syst)}~\gev$.
The combination between ME and NW measurements is performed after
assessing the statistical correlation which amounts to $(64\pm2)\%$~\cite{Abazov:2017ktz}.
The combined value is obtained using the BLUE method~\cite{Valassi:2003mu}.
It  constitutes the final D0 Run~II dilepton measurement with a relative uncertainty of 0.9\% below the percent level:
 $ m_t = 173.50\pm 1.31  \mbox{ (stat)}   \pm 0.84\mbox{ (syst)}~\gev
$. 

\subsection{Pole mass from cross section}
The technique exploits the dependence of the theoretical prediction of the
$\ttbar$ cross section as a function of the top quark pole mass.
D0 performs an extraction using the  measurement of the inclusive $\ttbar$ cross-section based on the full Run~II dataset~\cite{massxs}. In this analysis, the top quark mass is extracted  accounting
for the dependence of the measurement as a function of the mass due to acceptance effects and using the
NNLO cross-section calculation of  top++~\cite{top++}. D0 achieves a 1.9\% precision:
$m_t=172.8\pm 1.1 \mbox{ (theory)} ^{+3.3}_{-3.1} \mbox{ (experimental)}~\gev$.

More recently, D0 performs a similar extraction using  its measured  differential $\ttbar$ cross section~\cite{Abazov:2014vga} as a function of sensitive kinematic variables: the mass of the $t\bar t$ system ($m(\ttbar)$) and the the transverse momentum of the top quarks ($\pt(t/\bar t)$).
The mass is extracted from a fit of the NLO and NNLO cross sections~\cite{Czakon:2016ckf} to the unfolded D0 data. The mass is extracted with different factorization and renormalization scale and PDF to obtain the theory uncertainty.
The measurements using NNLO calculations is
$m_t=169.1 \pm 1.1 \mbox{ (theory)} \pm 2.2 \mbox{ (experimental)}~\gev$ which achieves
a 1.5\% accuracy~\cite{D0conf6473}.
This represents a 25\% improvement over the measurement using the inclusive cross-section, demonstrating the power of the method.

\subsection{D0 combination}

\begin{figure}[h]\center
\includegraphics[width=0.8\textwidth]{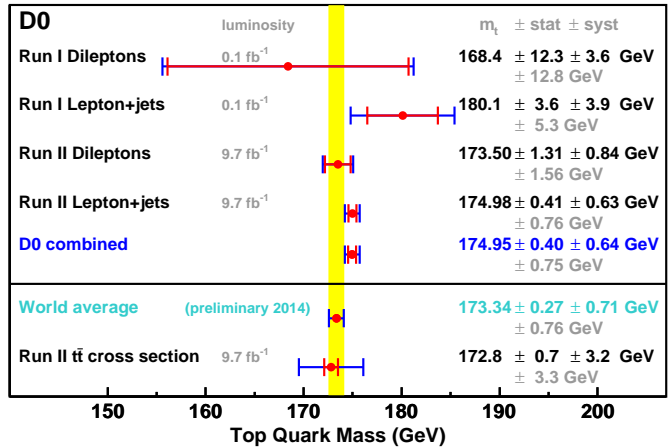}    
\caption{Summary of D0 top quark mass measurements and their combination. \label{fig:d0mass}}
\end{figure}

D0 has recently produced its legacy combined result regarding the direct measurement of the top quark mass~\cite{Abazov:2017ktz}.
 The combination  is based on measurements in the lepton+jets and dilepton channels,
 using all the data collected in  Run~I and  Run~II,
corresponding to integrated luminosities of 0.1~\invfb\ and 9.7~\invfb, respectively. The measurement of the pole mass of the top quark based on cross-sections is not included.
The combined mass, calculated with the BLUE method and shown in Fig.~\ref{fig:d0mass}, is 
$m_t=174.95\pm 0.40 \mbox{ (stat)} \pm 0.64 \mbox{ (syst)}=174.95\pm0.75~\GeV$, which has a 0.43\% relative uncertainty.
Such an accuracy is much better than what had been anticipated at the beginning of Run~II.
The dominant uncertainty sources  are the statistical uncertainty, the JES uncertainty arising from the lepton+jets calibration (0.41~\gev, also of statistical origin), and the signal modeling (0.35~\gev).

\subsection{Tevatron combination}
\begin{figure}[h]\center
\includegraphics[width=0.6\textwidth]{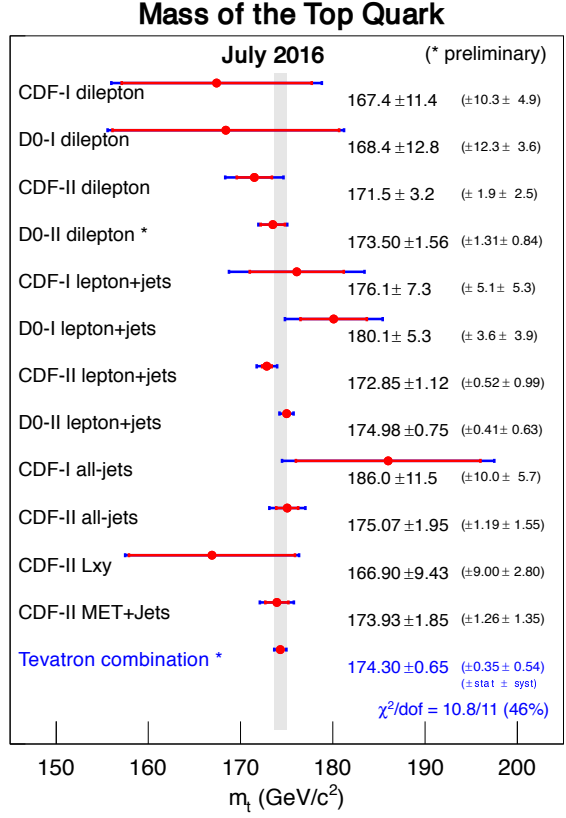}  
\caption{Summary of D0 and CDF top quark mass measurements and their combination. \label{fig:tevmass}}
\end{figure}

The combined Tevatron top quark mass  is using the most recent published
mass measurements of both D0 and CDF. All uncertainties and their correlations are accounted for using the BLUE method.
As for the D0 measurement, the dominant sources of uncertainty are the statistical uncertainty, the JES uncertainty arising from the lepton+jets calibration (0.31~\gev), and the signal modeling (0.36~\gev).
The combined mass is
$m_t=174.30\pm 0.35\mbox{ (stat)} \pm 0.54 \mbox{ (syst)}~\GeV$, which has a 0.37\% relative uncertainty~\cite{Tevatron:2016combo}.
The summary of  Tevatron measurements is presented in Fig.~\ref{fig:tevmass}.

\section{Forward-backward asymmetry in $\ttbar$ production}

\def\yt {\ensuremath{ y_t}}
\def\ytbar {\ensuremath{ y_{\bar t}}}
\def\dyttbar{\ensuremath{\Delta y_{\ttbar}}}
\newcommand{\defAll}    {\ensuremath{\All = \frac{N(\Delta\eta > 0) - N(\Delta\eta < 0)}{N(\Delta\eta > 0) + N(\Delta\eta < 0)}}}
\newcommand{\Alfb}      {\ensuremath{A^{\ell}_{\rm FB}}}
\newcommand{\defAlfb}   {\ensuremath{\Alfb = \frac{N_\ell(q\times \eta>0) - N_\ell(q\times \eta<0)}{N_\ell(q\times \eta>0) + N_\ell(q\times \eta<0)}}}
\newcommand{\Attfb}       {\ensuremath{A^{t\bar t}_{\rm FB}}}
\newcommand{\Att}       {\Attfb}
\newcommand{\defAtt}{\ensuremath{\Att=   \frac {  N ( \dyttbar>0) -  N ( \dyttbar <0) }{  N ( \dyttbar>0) + N ( \dyttbar <0) }}}
\newcommand{\All}       {\ensuremath{A^{\ell\ell}_{\rm FB}}}

\begin{figure}[h]\center
\includegraphics[width=0.6\textwidth]{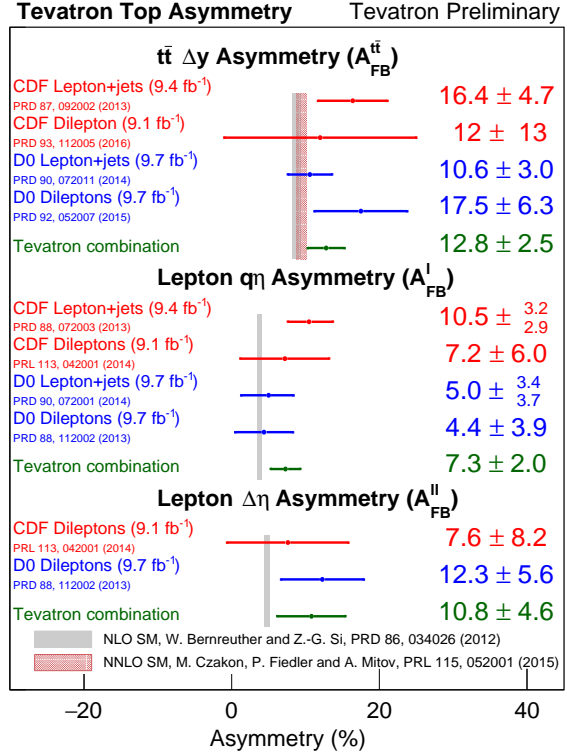}     
\caption{Summary of Tevatron  $p\bar p\to t\bar t$ asymmetry measurements and their combinations. \label{fig:tevafb}}
\end{figure}

The theory of strong (QCD) and electroweak (EW) interactions predicts a positive forward-backward asymmetry in the $\ppbar\to\ttbar$ production: the (anti)top quarks tend to go in the same direction as the  incoming (anti)protons.
When the measurements were first performed at Tevatron Run~II, a large departure from the SM expectations was observed which brought a
lot of excitement~\cite{Abazov:2007ab,Aaltonen:2008hc,Aaltonen:2011kc,Abazov:2011rq}.
Since then,  the theory prediction has been improved by including higher order QCD and EW corrections~\cite{Czakon:2014xsa},
and both CDF and D0 have completed their asymmetry measurement program using dilepton and lepton+jets channels and analyzing the full Run~II dataset.
They now provides the combination of their measurements obtained with the BLUE method, accounting for all uncertainties and their correlations~\cite{TevconfAFB}.
As the different measurements are dominated by the uncorrelated statistical uncertainties,
the overall correlations are small, at the level of 10\%, mainly due to signal modeling uncertainty.
The combination is performed for the three asymmetries,  based either on the reconstructed rapidities of the top and antitop ($\Delta y=y_t-y_{\bar t}$), or on the lepton pseudorapidities ($\eta$), or on the difference between lepton rapidities for dilepton channels ($\Delta\eta=\eta_{\ell^+}-\eta_{\ell^-}$).
Figure~\ref{fig:tevafb} summarizes the measurements for the three asymmetries and their combinations,
\defAtt, \defAlfb, and \defAll, where each $N$ represents a number of events corrected for acceptance and efficiency effects.
All combined asymmetries are compatible within 1.3--1.6 standard deviation to the NNLO predictions of the SM. 

\section{CP violation in $B^+\to \mu^+\nu\bar  D^0$ decay}
We defined a CP violating asymmetry in the decay of charge $B$ meson  as:
$A^{\mu D^0}=
\frac{\Gamma( B^- \to \mu^-\bar{\nu}_\mu D^0) - \Gamma( B^+ \to \mu^+{\nu_\mu} \bar D^0)}
{\Gamma( B^- \to \mu^-\bar{\nu}_\mu D^0) +\Gamma( B^+ \to \mu^+{\nu_\mu} \bar D^0)}
$.
A non-zero asymmetry of
$A^{\mu D^0}\approx 0.8\%$ would explain why the like-sign dimuon asymmetry  measured at D0~\cite{Abazov:2013uma} shows
a deviation from the SM at the 3.6 standard deviation level. To measure $A^{\mu D^0}$, we select a sample of  $D^0\to K^-\pi^+$ candidates in the mass range $1.40<M(K\pi)<2.20\ \GeV$ in data recorded by  single muon triggers in 10.4~\invfb of \ppbar\ collisions.
A simultaneous fit of the sum and difference of the $B^+$ and $B^-$ spectrum allows to disentangle the signal from the background to measure the raw asymmetry.
Once the  effects of detector asymmetries, and of known asymmetries from physics sources are subtracted, we correct for the dilution
arising from non-$B^+$ sources of $\mu D^0$ events and obtain $A^{\mu D^0}=[-0.14\pm0.14\mbox{ (stat)}\pm 0.14\mbox{ (syst)}]\%$~\cite{Abazov:2016kfr}, which is in agreement with the SM expectation of 0. This is the first time this measurement is performed on a collider. 

\section{New state X(5568) decaying to $B^0_s\pi^{\pm}$}
\begin{figure}[h]\center
\includegraphics[width=0.6\textwidth]{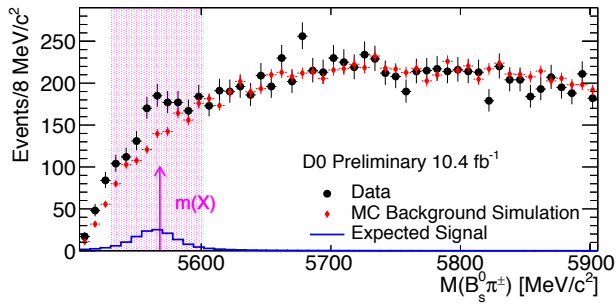}     \caption{The $m(B^0_s\pi^{\pm})$ distribution in the semi-leptonic channel, together with a background simulation and the expected signal.\label{fig:BspiSL}}
\end{figure}
Last year D0 published the first evidence for a new state decaying to $B^0_s\pi^{\pm}$~\cite{D0:2016mwd}, which was observed  at the 5.1 or 3.9 standard deviation level, depending on the selection requirements.
The signal is seen in the hadronic decay of the $B^0_s$ meson: $B^0_s\to J/\psi\phi$, $J/\psi\to \mu^+\mu^-$, $\phi \to K^+K^-$.
Its mass and natural width are measured as $M_X=5567.8\pm 2.9$~MeV and $\Gamma_x=21.9\pm6.4$~MeV.
The relatively large width indicates the decay cannot be mediated by the  weak (flavor changing) interaction, so as the 4 flavors of the final state ($u$, $d$, $b$, and $s$)
should also be present in the initial state.
Therefore this might be the first evidence for a tetraquark made of 4 different flavors.
Since the first claim,  LHC experiments, LHCb~\cite{Aaij:2016iev} and CMS~\cite{CMS_Bspi},
performed similar analyses but did not observed any significant signal.
A mystery remains at the moment, why the D0 observation could not  be  confirmed by LHC data.

To  confirm its signal, D0 is now looking at a different  channel involving the semi-leptonic decays of the $B^0_s$ mesons:
$B^0_s\to D_s^- \mu^+\nu_{\mu}$,  $D_s^- \to\phi\pi^- $ and $ \phi\to K^+K^-$.
The data are selected by searching for a $D_s$ peak, associated with the presence of a muon, employing similar kinematic selection as in the hadronic channel. The fraction of $D_s\mu$ candidates arising from real  $B^0_s$ events is estimated to 90\%, so as the sample consists of $\approx 8000 $ $B^0_s$ and approximately the same amount of combinatorial background. Extrapolating from the yield measured in the hadronic channel, one expects to have $\approx 150$ $X(5568)$ candidates in this sample.
For Summer 2016 conferences, D0 released the invariant mass spectrum   shown in Fig.~\ref{fig:BspiSL}, comparing data, a background simulation, and the expected signal~\cite{D0conf6488}.
No quantitative estimate of the signal yield or of the signal significance was made, as the modeling of the background was not settled at the time. However, the presence of the $X$ signal at the right place is qualitatively confirmed in Fig.~\ref{fig:BspiSL}.

Shortly after La Thuile conference, D0  released the preliminary quantitative confirmation of the $X(5568)$ state in the semi-leptonic channel, with a significance of 3.2 standard deviation~\cite{D0conf6496}.

\section{Conclusion}
Five years after the shutdown of Tevatron, new measurements and  new results are appearing  from the Tevatron experiments.
In the realm of top quark physics, we provide precise measurements of the top quark mass, at  0.43\% accuracy level at D0, and at  0.37\% accuracy level at Tevatron.
We also provides precise measurements of the $\ttbar$ forward-backward asymmetries which are in agreement with the Standard Model.
A new CP violating asymmetry is measured to be compatible with 0, using the semileptonic $B^+$ decays  at D0.
A new QCD state $X(5568)$ decaying to $B_s^0\pi^{\pm}$ is observed using  hadronic decays of the $B_s^0$ mesons at D0, and is also confirmed in the semileptonic decay channels.
See Ref.~\cite{D} for more recent results arising from Tevatron.

\bibliographystyle{varenna.bst}

\end{document}